\documentclass[journal]{IEEEtran}
\usepackage{amsmath,amssymb}
\usepackage{multirow}
\usepackage{graphicx}
\usepackage{bm}
\usepackage{algorithm}
\usepackage{algorithmic}
 \usepackage{amsmath}
 \usepackage{booktabs}
 \usepackage{graphicx}
 \usepackage{hyperref}
 \usepackage{wasysym}
 \usepackage{multirow}
 \usepackage{booktabs}
 \usepackage{pifont}
 \usepackage{setspace}
 \usepackage[table]{xcolor}
\newcommand{\myparagraph}[1]{\vspace{6pt}\noindent\textbf{#1.}\quad}

\usepackage{array}
\usepackage[caption=false,font=normalsize,labelfont=sf,textfont=sf]{subfig}
\usepackage{textcomp}
\usepackage{stfloats}
\usepackage{url}
\usepackage{verbatim}
\usepackage{cite}
\usepackage{setspace}

\usepackage{amsthm}

\begin{document}

\title{Conflict-Aware Meta-Learning for Generalizable Speech Deepfake Detection}

\author{Siqing Qin,  
        Kong Aik Lee*\thanks{* Corresponding author},
        Youzhi TU,
        Eng Siong Chng,
        and Man-Wai Mak
\thanks{Siqing Qin,  
        Kong Aik Lee,
        Youzhi TU, and Man-Wai Mak are with the Department of  Electrical and Electronic Engineering, The Hong Kong Polytechnic University.}
\thanks{Eng Siong Chng is with the College of Computing and Data Science, Nanyang Technological University.}
}


\maketitle

\begin{abstract}
Speech deepfake detectors built on large self-supervised learning front-ends achieve high in-domain accuracy, yet suffer degradation under domain shift caused by unseen synthesizers, transmission codecs, or unfamiliar acoustic conditions. Meta-learning, particularly \textit{model-agnostic meta-learning for domain generalization} (MLDG), addresses this issue by simulating domain shifts through episodic meta-train and meta-test splits. However, applying MLDG directly introduces a bi-level optimization challenge that remains unsolved. First, effective domain-shift emulation requires genuine distributional gaps between meta-splits, yet such gaps can frequently produce gradient conflicts between the meta-train and meta-test objectives, destabilizing the bi-level optimization. Second, the stochastic nature of mini-batch gradients causes the reference direction to fluctuate across iterations, resulting in an optimization trajectory that fails to achieve domain generalization. We propose a \textit{conflict-aware meta-learning} (CAML) framework, which addresses both challenges through three complementary components. Firstly, base \textit{domain gradient surgery} (base-DGS) introduces an asymmetric projection that treats the meta-train gradient as a primary anchor and removes only the conflicting component of the meta-test gradient. Secondly, EMA-DGS replaces the unstable anchor with its exponential moving average (EMA), providing a stable optimization trajectory anchor that substantially reduces domain-shift-induced conflict detections. Thirdly, layer-wise DGS (LW-DGS) further extends the framework to individual parameter tensors, enabling fine-grained, layer-specific intervention that targets conflict-prone layers while leaving well-aligned parameters intact. Experimental results demonstrate that the proposed CAML framework achieves lower EER than the compared baselines on most cross-dataset evaluations across multiple cross-dataset benchmarks and architectures, achieving a maximum relative improvement of 45.2\% in terms of EER.
\end{abstract}


\section{Introduction}
\label{sec:intro}
\IEEEPARstart{S}{peech} deepfakes are synthesized or manipulated speech to deceive human listeners and authentication systems. This poses a threat to security and privacy \cite{asvspoof2019, yamagishi2021asvspoof, asvspoof5, cfad2024}. Due to advances in deep generative models, synthesized voices have become increasingly difficult to distinguish from genuine speech. Early detection methods relied on conventional hand-crafted features (such as LFCCs or CQCCs) \cite{yang2018extended, yang2020long, choi2022overlapped, li2022long, li2023investigation} and were combined with back-end classifiers \cite{lavrentyeva2019stc, li2022role, rawbmamba}. State-of-the-art detectors now frequently leverage large \textit{self-supervised learning} (SSL) front-ends (e.g., XLS-R \cite{babu2021xls}, WavLM \cite{chen2022wavlm})  to extract rich acoustic representations directly from raw waveforms. These SSL front-ends achieve high accuracy under controlled experimental settings \cite{lv2022fake, tak2022wav2vec, phukan2024heterogeneity, xiao2025xlsr, li2026dure}.

However, current speech deepfake detectors exhibit performance degradation when deployed in real-world scenarios. This vulnerability stems from \textit{domain shift} \cite{li2025survey, dg2024, pascu2025easy} due to the disparity between training data and unseen test environments. Such shifts encompass new speech synthesis algorithms and attack types, varying transmission codecs, or unfamiliar acoustic backgrounds. While unsupervised domain adaptation (UDA) seeks to align feature distributions using unlabeled target data, it may be impractical when target data are inaccessible due to privacy or operational constraints \cite{chen2025continual, 2025conti2}. Consequently, \textit{domain generalization }(DG), which aims to learn domain-invariant representations solely from available source domains, has emerged as a practical paradigm for defending against deepfake attacks \cite{huang25e_interspeech, ren2025dg}.

To achieve DG, the standard baseline is empirical risk minimization (ERM) \cite{vapnik1991erm}, in which data from all accessible source domains are pooled for naive joint training. Although it is robust when large amounts of diverse data are available, ERM fails to exploit the inherent structural differences between domains explicitly. Recent studies demonstrate that meta-learning techniques, which simulate domain shifts during training, can surpass ERM \cite{2025metageneralizable, metatrim2024}. The meta-learning framework MLDG \cite{li2018mldg} has emerged as a promising alternative. Extensions such as multi-task meta-learning (MTML) have been proposed to stabilize the meta-optimization trajectory and help escape local minima \cite{metatrim2024}. Furthermore, MLDG has been combined with parameter-efficient tuning methods such as low-rank adaptation (LoRA) \cite{hu2022lora}, with these approaches outperforming ERM in cross-domain evaluations \cite{2025metageneralizable}. 

Applying MLDG to large-scale SSL backbones reveals an optimization dilemma. As shown in our case study (Section~\ref{sec:case_study}), effective domain-shift emulation requires a substantial gap between meta-train and meta-test domains, but such gaps often produce conflicting gradients, with negative cosine similarity in over 43\% of training iterations. Naive gradient aggregation can then interfere with optimization despite providing informative domain shifts.

Traditional gradient surgery methods, such as PCGrad \cite{yu2020gradient}, resolve conflicts through symmetric projection in multi-task learning and speech processing \cite{truong2025addressing}. However, meta-train and meta-test gradients have asymmetric roles in MLDG. The former provides the primary source-domain learning signal, whereas the latter encourages generalization. Symmetric projection may thus weaken source-domain adaptation. A second challenge is the stochastic volatility of mini-batch gradients, which can cause inconsistent conflict detection and unnecessary projections. Motivated by EMA mechanisms in self-supervised learning \cite{morales2024exponential,cai2021exponential}, optimization \cite{kingma2014adam}, and gradient estimation \cite{Li2024fsam}, we use an exponential moving average of meta-train gradients to provide a stable projection anchor.

\begin{figure*}
    \centering
    \includegraphics[width=1.0\linewidth]{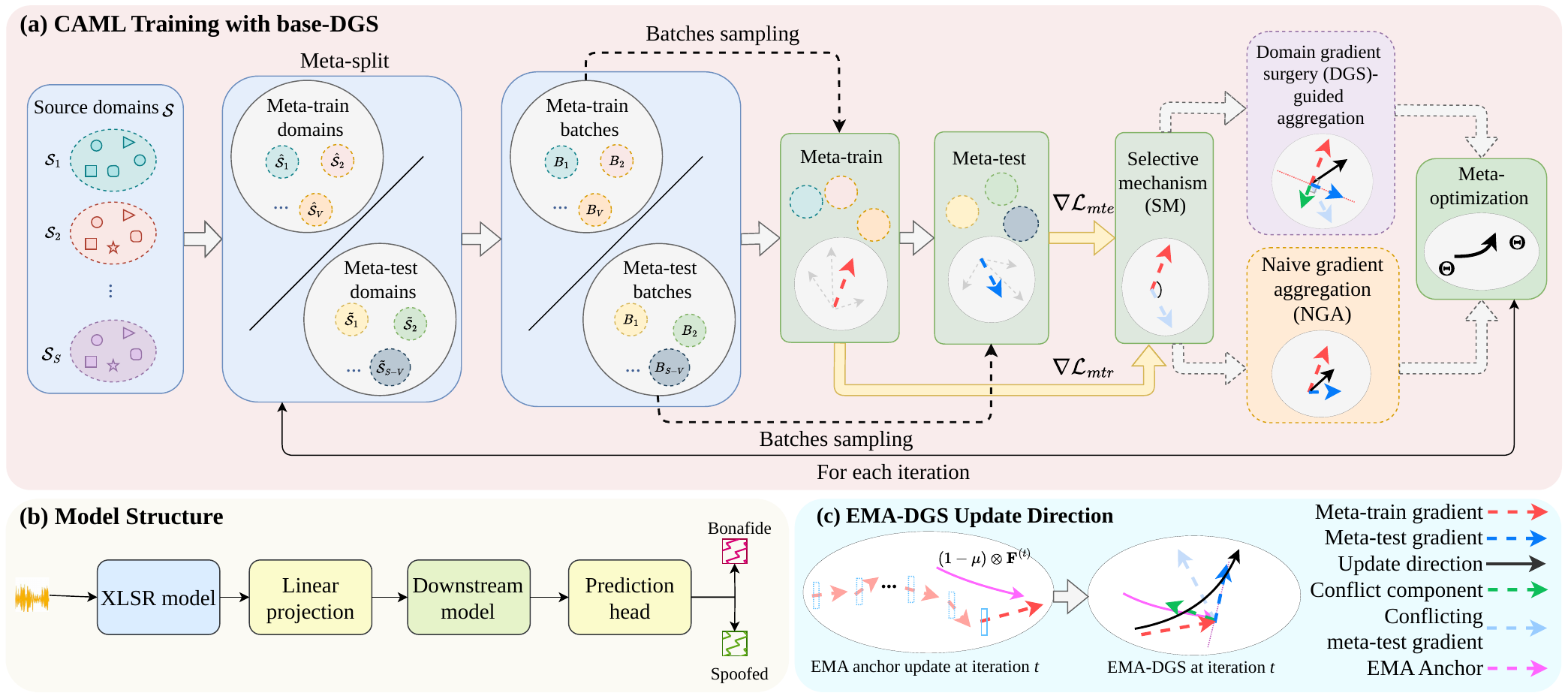}
    \caption{Overview of the proposed CAML framework. (a) Each training iteration partitions source domains into meta-train and meta-test splits to emulate domain shift. The meta-train gradient $\bm{F}$ and meta-test gradient $\bm{G}$ are passed to a selective mechanism (SM) that detects conflicts ($\langle\bm{F},\bm{G}\rangle<0$) and applies asymmetric Domain Gradient Surgery (DGS): globally as a single projection (base-DGS) or independently per parameter tensor (LW-DGS). (b) The detection backbone consists of an SSL front-end, a linear projection, a downstream temporal model, and a binary classification head. (c) EMA-DGS replaces the volatile per-batch anchor with its exponential moving average $\bm{F}_{\text{EMA}}$ (pink arrow). The EMA anchor provides a stable reference trajectory to guide surgery when meta-test gradients conflict.}
    \label{fig:modeloverview}
\end{figure*}

In our previous study \cite{qin2026dgsmldg}, we proposed a domain gradient surgery-guided MLDG (DGS-MLDG) method using asymmetric gradient projection to mitigate gradient conflicts, along with its layer-wise variant (LW-DGS-MLDG). Building upon this work, we propose the \textit{conflict-aware meta-learning framework} (CAML), which integrates asymmetric surgery with temporally stable conflict detection at global and layer-wise levels. The framework comprises three components (Figure~\ref{fig:modeloverview}). First, \textit{base domain gradient surgery} (base-DGS) adopts the asymmetric projection of DGS-MLDG, preserving the meta-train gradient as an anchor while correcting only the conflicting component of the meta-test gradient. Second, we propose \textit{EMA-DGS}, which replaces the instantaneous meta-train gradient with its exponential moving average (EMA) as the anchor for conflict detection and projection, improving temporal stability. Third, we extend our previous layer-wise DGS (LW-DGS)\footnote{In this paper, we use the term ``layer'' to refer to an individual parameter tensor, such as a bias vector or a weight matrix.} with an EMA-enhanced variant (LW-EMA-DGS). Both variants apply surgery per parameter tensor, while LW-EMA-DGS additionally uses EMA anchors for stable, targeted intervention without altering well-aligned tensors.

The primary contributions of this work are summarized as follows:
\begin{itemize}
    \item We conduct a systematic case study identifying the dilemma in MLDG: effective domain-shift emulation improves generalization but often induces conflicting gradients.
    \item Building on our previous asymmetric DGS, we introduce EMA-DGS to stabilize conflict detection with an EMA meta-train anchor and provide a variance-reduction analysis (Section~\ref{subsec:ema_dgs}).
    \item We extend the previously proposed LW-DGS with EMA-based, tensor-level conflict detection and projection (LW-EMA-DGS), enabling stable intervention at conflict-prone parameter tensors.
    \item We conduct evaluations across six zero-shot benchmarks and three architectures, demonstrating the effectiveness and scalability of CAML for speech deepfake detection.
\end{itemize}

\section{Related Work}
\label{sec:related}

This section reviews domain generalization for speech deepfake detection, meta-learning approaches, and gradient surgery methods.

\subsection{Generalization in Speech Deepfake Detection}
Previous detectors use hand-crafted time-frequency representations \cite{tak2020explainability,davis1980comparison} to detect spoofing artifacts, but often generalize poorly to unseen TTS methods. This limitation motivated raw-waveform and self-supervised learning (SSL) detectors, including systems using Wav2Vec 2.0 \cite{baevski2020wav2vec}, which extract contextual representations learned from unlabeled speech.

However, newly emerging attacks remain challenging for SSL-based detectors. Domain generalization methods include extensive data augmentation, such as RawBoost \cite{rawboost}, to simulate channel and noise disturbances, and feature disentanglement to emphasize spoof-related cues. Augmentation may not anticipate fundamentally new synthesis algorithms, while disentanglement can require multi-component architectures. Meta-learning provides an alternative for learning representations robust to domain shifts.

\subsection{Meta-learning for Domain Generalization}
\label{subsec:meta}
Meta-learning, or ``learning to learn,'' aims to maximize model adaptability \cite{finn2017model}. MLDG \cite{li2018mldg} adapts this principle to domain robustness. In speech deepfake detection, meta-learning has been combined with silence-trimming augmentation \cite{metatrim2024}, while other work integrates MLDG with parameter-efficient LoRA tuning \cite{2025metageneralizable,hu2022lora}. However, these approaches generally use linear aggregation of meta-train and meta-test gradients, which can be problematic under substantial domain shifts or in-the-wild settings.

\subsection{Gradient Surgery}
In multi-task learning, a \textit{gradient conflict} occurs when two objectives have a negative gradient inner product, i.e., $\langle g_i,g_j\rangle<0$ \cite{yu2020gradient}. PCGrad \cite{yu2020gradient} and CAGrad \cite{liu2021conflict} address such interference, and PCGrad has also been applied to speech deepfake detection with dual-pass augmentation \cite{truong2025addressing}. However, symmetric treatment may be poorly matched to MLDG's bi-level hierarchy. The meta-train gradient $\bm{F}$ represents source-domain learning, whereas the meta-test gradient $\bm{G}$ provides a generalization signal. Projecting both can weaken the primary source-domain update. It motivates us to perform asymmetric gradient surgery.

\section{Case Study on Domain Shift in Meta-Learning}
\label{sec:case_study}

This section describes an empirical study that establishes the motivation of our work. It demonstrates that not all attack-type variations constitute a genuine domain shift, justifying the need for a principled domain construction strategy. Additionally, this section shows that well-constructed meta-learning splits may suffer from persistent gradient conflicts, which motivates the asymmetric gradient surgery in Section~\ref{sec:meth}.

While meta-learning frameworks aim to mitigate domain shift by simulating domain shifts during training, a process we term \textit{domain shift emulation} (DSE), the effectiveness of this simulation depends heavily on how domains are constructed. 
Specifically, does simply splitting data by arbitrary attack types guarantee a meaningful domain shift? And how can we structure the meta-learning process to maximize the probability of encountering effective shifts? To address these questions, we conducted a two-stage pilot study using an SSL-pretrained XLSR backbone.

\subsection{Stage 1: Attack Type Variation vs. Domain Shift}
First, we investigate whether all variations in attack types constitute a meaningful domain shift that challenges the deepfake detector. We utilize attack types from the ASVspoof 2019 LA training set (Attacks 1--6) and the ASVspoof 5 training set (Attacks 7--14). The model is trained on a single attack type and evaluated against all other attack types to measure cross-domain performance.

\begin{figure}[t]
    \centering
    \includegraphics[width=1.0\linewidth]{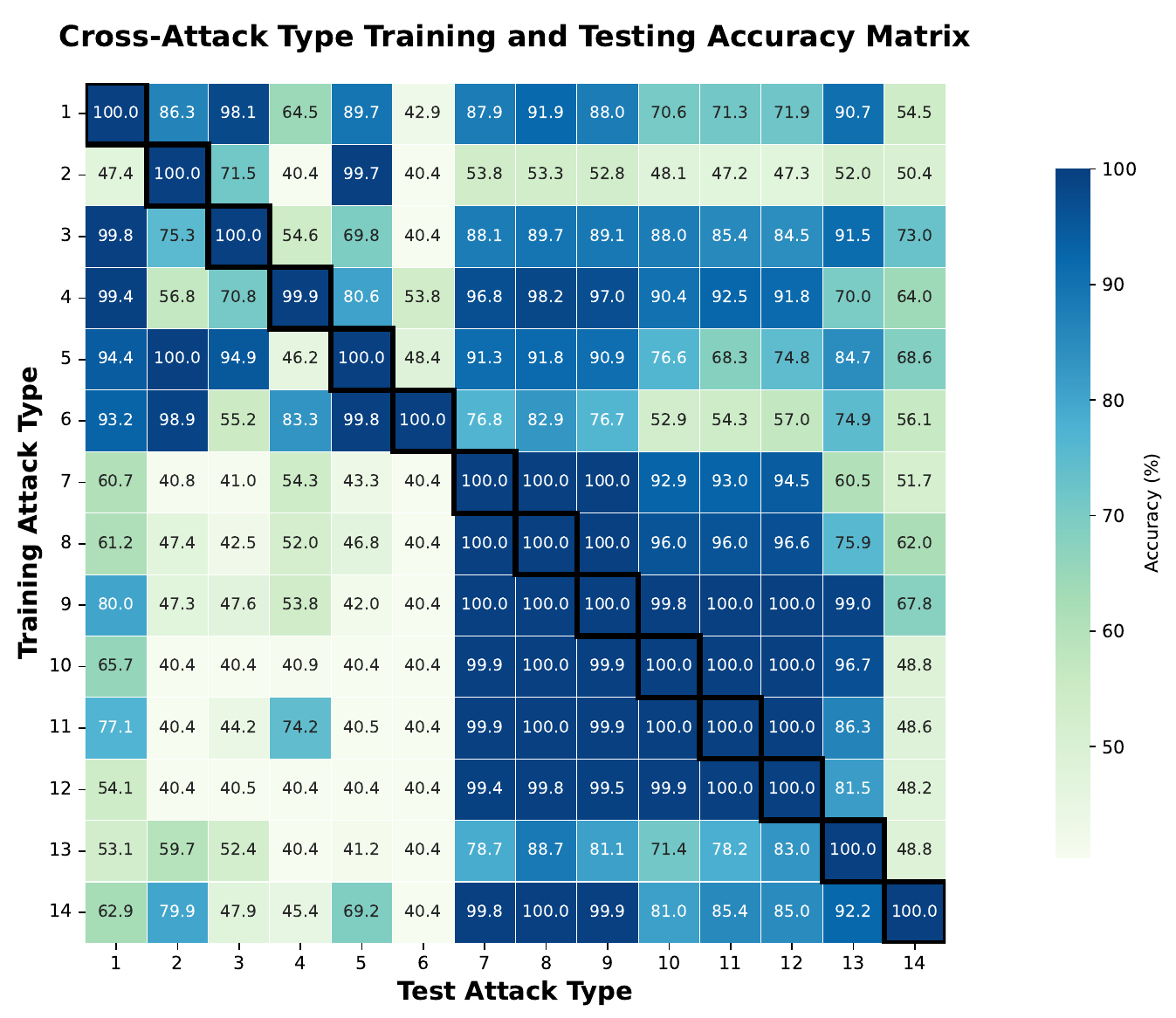}
    \caption{Heatmap of accuracy under cross-attack evaluation. Varied performance drops indicate that not all changes in attack types result in effective domain shifts.}
    \label{fig:case1}
\end{figure}

As illustrated in Figure \ref{fig:case1}, we observe significant heterogeneity in generalization performance. Shifts between certain attack types (e.g., within the ASVspoof 5 subset) often yield relatively high accuracy, suggesting that these attacks share similar acoustic artifacts and do not constitute a sufficient domain shift to challenge the model. Conversely, shifts between distinct datasets (e.g., from Attack 1--6 to 7--14) result in large performance drops, with accuracy decreasing to 40.4\%. 

\textit{Insight 1:} Relying on random, single-pair domain splits risks sampling two highly similar domains, yielding weakly informative meta-gradients that contribute no generalization cues. Therefore, to ensure robust optimization, we must adopt a multi-domain aggregation strategy at each meta-update to statistically guarantee the observation of DSE.

We monitored training under two distinct settings:
\begin{itemize}
    \item \textbf{Without DSE}: Meta-train (Attacks 7-9) and Meta-test (Attacks 10-12) lack a fundamental shift.
    \item \textbf{With DSE}: Meta-train (Attacks 7-9) and Meta-test (Attacks 4-6) possess a severe distribution gap.
\end{itemize}

When evaluated on an entirely unseen dev set (ASVspoof 2019 LA), the ``Without DSE" model failed to generalize (67.08\% accuracy), whereas the ``With DSE" model performed robustly (95.04\%). This confirms that simulated shifts are critical for meta-learning.

\subsection{Stage 2: The Dilemma of Gradient Conflicts}
\begin{figure}[t]
    \centering
    \includegraphics[width=1.0\linewidth]{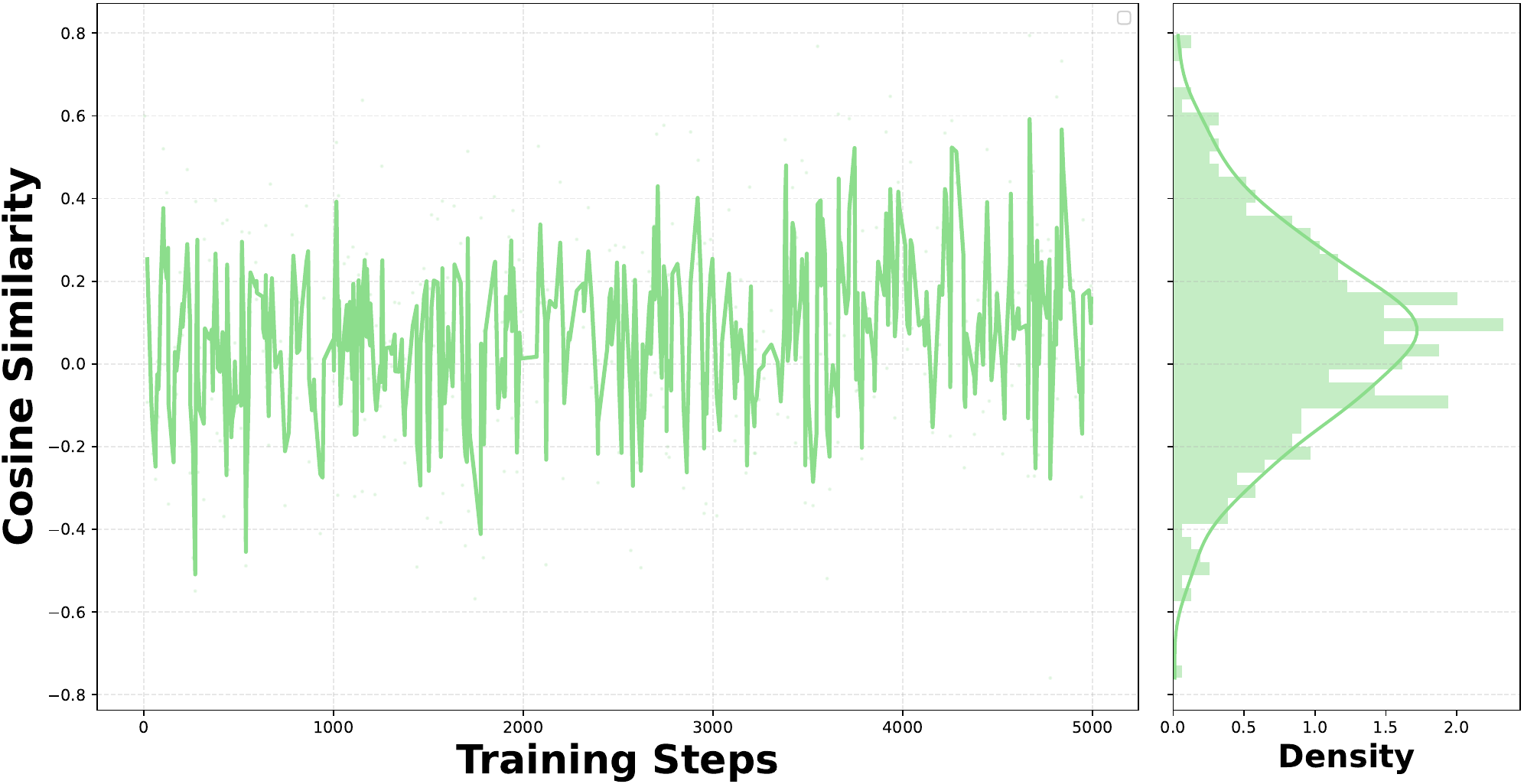}
    \caption{Cosine similarity between meta-train and meta-test gradients with effective domain shift emulation (DSE). Strong shifts trigger negative similarities (conflicts).}
    \label{fig:coscase_placeholder}
\end{figure}

If we successfully simulate a domain shift via multi-domain splits,
the cosine similarity between the meta-train gradient $\bm{F}$ and meta-test gradient $\bm{G}$ may fluctuate significantly at each step. As depicted in Figure \ref{fig:coscase_placeholder}, the effective DSE setting causes frequent negative cosine similarities $\langle \bm{G}, \bm{F} \rangle < 0$.

\textit{Insight 2:} Real generalization learning in MLDG is often accompanied by gradient conflicts. When $\bm{F}$ and $\bm{G}$ conflict with each other, naive gradient aggregation (NGA) $\bm{F} + \beta\bm{G}$ can reduce the gradient descent component for one or both objectives and may destabilize optimization. Therefore, we argue that meta-learning needs DSE to force generalization. Simultaneously, we need to deal with the harmful conflicts that lead to unstable optimization.

\section{Methodology}
\label{sec:meth}

Building on the empirical findings in Section~\ref{sec:case_study}, this section presents the CAML framework. We begin with MLDG preliminaries and an analysis of implicit gradient interactions. We then introduce the three core components in order of increasing sophistication: base-DGS and EMA-DGS, which stabilize conflict detection via an exponential moving average anchor; and LW-DGS, which extends the framework to per-layer granularity for fine-grained intervention.


\subsection{Preliminaries}
\label{subsec:prelim_mldg}
As illustrated in Figure \ref{fig:modeloverview}(a), MLDG involves a meta-train (source adaptation) phase and a meta-test (generalization) phase. The source domains $\mathcal{S}={\mathcal{S}_i}|_{i=1}^{S}$ are partitioned into $V$ meta-train domains $\hat{\mathcal{S}}$ and $S-V$ meta-test domains $\tilde{\mathcal{S}}$ in each iteration. The meta-train loss $\mathcal{L}_{\text{mtr}}$ guides the inner-loop adaptation, while the meta-test loss $\mathcal{L}_{\text{mte}}$ evaluates the generalization on the adapted parameters $\bm{\Theta}' = \bm{\Theta} - \alpha \nabla_{\bm{\Theta}} \mathcal{L}_{\text{mtr}}(\bm{\Theta})$. The overall objective minimizes the weighted sum:
\begin{align}
    \mathop{\arg\min}_{\bm{\Theta}}\mathcal{L}_{\text{mtr}}(\bm{\Theta}) + \beta\mathcal{L}_{\text{mte}}(\bm{\Theta} - \alpha \nabla_{\bm{\Theta}} \mathcal{L}_{\text{mtr}}(\bm{\Theta})),
    \label{equ:mldg}
\end{align}
where $\beta$ weights the meta-test phase. 
Standard MLDG uses naive gradient aggregation (NGA) to linearly combine gradients. NGA aims to help the model perform well on source domains while maintaining the capacity for fast adaptation to unseen domains. 

\subsection{Analysis of MLDG's Implicit Gradient Interaction}
\label{subsec:gradient_analysis}
\label{subsubsec:mldg_taylor_analysis}

To establish a theoretical foundation for our proposed refinement, we analyze how standard MLDG implicitly optimizes the relationship between meta-train and meta-test splits. We perform a second-order Taylor expansion on the meta-test loss $\mathcal{L}_{\text{mte}}$ around the initial parameters $\bm{\Theta}$. Let $\bm{F} = \nabla_{\bm{\Theta}} \mathcal{L}_{\text{mtr}}(\bm{\Theta})$ denote the meta-train gradient, $\bm{G} = \nabla_{\bm{\Theta}} \mathcal{L}_{\text{mte}}(\bm{\Theta})$ denote the meta-test gradient, and $\bm{H}_{\text{mte}}$ denote the Hessian matrix of the meta-test loss at $\bm{\Theta}$. 

The inner update shifts the parameters by $-\alpha \bm{F}$ to yield adapted parameters $\bm{\Theta}'$. The meta-test loss at $\bm{\Theta}'$ can be approximated as:
\begin{align}
    \mathcal{L}_{\text{mte}}(\bm{\Theta} - \alpha \bm{F}) &\approx \mathcal{L}_{\text{mte}}(\bm{\Theta}) - \alpha \langle \bm{F}, \bm{G} \rangle + \frac{1}{2} \alpha^2 \bm{F}^{T} \bm{H}_{\text{mte}} \bm{F}.
    \label{eq:mldg_taylor_2nd_order}
\end{align}
The overall objective of MLDG is to minimize the sum $\mathcal{L}_{\text{mtr}}(\bm{\Theta}) + \beta \mathcal{L}_{\text{mte}}(\bm{\Theta}')$. By examining (\ref{eq:mldg_taylor_2nd_order}), it becomes evident that to minimize the first-order interaction term ($-\alpha \langle \bm{F}, \bm{G} \rangle$), standard MLDG inherently encourages the maximization of the inner product $\langle \bm{F}, \bm{G} \rangle$. 

This implicit mechanism exposes a critical vulnerability when simulating severe domain shifts. When the meta-train and meta-test domains exhibit highly divergent acoustic characteristics, their optimization trajectories often diverge, leading to a negative inner product ($\langle \bm{F}, \bm{G} \rangle < 0$), which hinders the convergence speed or leads to a sub-optimal region. 
This mathematical consequence explains the destructive interference observed during naive gradient aggregation.

Furthermore, identifying and resolving this conflict is fundamentally complicated by the bi-level stochastic nature of MLDG. The gradients $\bm{F}$ and $\bm{G}$ are computed from small mini-batches, rendering them inherently noisy. Relying on an instantaneous, highly volatile vector $\bm{F}$ as a reference for detecting and correcting conflicts introduces severe temporal instability into the meta-optimization process. 

This analysis underscores the necessity of explicit gradient intervention (DGS) to neutralize destructive conflicts ($\langle \bm{F}, \bm{G} \rangle < 0$), and highlights the need to upgrade the intervention with temporal stabilization (EMA-DGS).
\subsection{Base-DGS: Domain Gradient Surgery}
\label{subsec:dgs}
\label{subsec:base_dgs}
Building upon the analysis, we first introduce the foundational mechanism of domain gradient surgery (DGS) to address the primary issue of destructive gradient conflicts. DGS employs a selective, asymmetric projection strategy, as shown in Figure \ref{fig:modeloverview}(a).

In meta-learning, the roles of meta-train and meta-test gradients are inherently asymmetric. The meta-train gradient $\bm{F}$ anchors the model's fundamental adaptation to the current batch, representing its core learning signal. Conversely, the meta-test gradient $\bm{G}$ serves as a regularizing signal, guiding the model towards generalizable solutions by simulating future shifts. Preserving $\bm{F}$ is therefore important to avoid compromising the immediate learning objective. Therefore, base-DGS declares $\bm{F}$ as the primary anchor, exclusively altering $\bm{G}$.

\textbf{Selective Mechanism (SM):} Base-DGS monitors the inner product between $\bm{F}$ and $\bm{G}$. If $\langle \bm{G}, \bm{F} \rangle \geq 0$, indicating constructive alignment or non-interference, the gradients are combined directly without intervention. However, if $\langle \bm{G}, \bm{F} \rangle < 0$, signifying destructive conflict as evidenced by our Taylor expansion, the DGS projection mechanism is enabled.

\textbf{Meta-Optimization with base-DGS:} When conflict is detected ($\langle \bm{G}, \bm{F} \rangle < 0$), base-DGS employs an asymmetric projection strategy. It treats $\bm{F}$ as the anchor for domain adaptation and removes only the detrimental component of $\bm{G}$ that directly opposes $\bm{F}$:
\begin{align}
\label{equ:base_dgs_proj}
\bm{G}_{\text{proj}} = 
\begin{cases}
    \bm{G} - \left(\frac{\langle \bm{G}, \bm{F} \rangle}{||\bm{F}||^2 + \epsilon} \right) \bm{F}, \quad &\text{if } \langle \bm{G}, \bm{F} \rangle < 0,\\
    \bm{G}, &\text{otherwise}.
\end{cases}
\end{align}
Here, $\epsilon$ is a small constant (e.g., $10^{-8}$) for numerical stability. This operation ensures that $\langle \bm{G}_{\text{proj}}, \bm{F} \rangle \geq 0$, thereby guaranteeing a conflict-free contribution from the meta-test gradient. The model parameters $\bm{\Theta}$ are then updated using the combined gradients:
\begin{equation}
\bm{\Theta} \leftarrow \bm{\Theta} - \gamma \left( \bm{F} + \beta \bm{G}_{\text{proj}} \right).
\label{eq:base_dgs_update}
\end{equation}

\begin{algorithm}[t]
\caption{Base-DGS-guided MLDG}
\label{alg:base_dgs}
\begin{algorithmic}[1]
\STATE \textbf{Input} Source domains $\mathcal{S}$, hyperparameters $\alpha, \beta, \gamma$, Mode $M \in \{\text{Global, LW}\}$
\STATE \textbf{Init} Model parameters $\bm{\Theta}$
\FOR{each iteration}
    \STATE \textbf{Split} $\hat{\mathcal{S}}$ and $\tilde{\mathcal{S}} \leftarrow \mathcal{S}$
    \STATE \textbf{Meta-train} 
    \STATE \quad Compute average gradient 
    \STATE \quad $\bm{F} = \frac{1}{|\hat{\mathcal{S}}|} \sum_{D_i \in \hat{\mathcal{S}}} \nabla_{\bm \Theta}\mathcal{L}_{\text{mtr}}(D_i, \bm{\Theta})$
    \STATE \textbf{Inner-update} $\bm{\Theta}' \leftarrow \bm{\Theta} - \alpha \bm{F}$
    \STATE \textbf{Meta-test} 
    \STATE \quad Compute average gradient 
    \STATE \quad $\bm{G} = \frac{1}{|\tilde{\mathcal{S}}|} \sum_{D_j \in \tilde{\mathcal{S}}} \nabla_{\bm \Theta} \mathcal{L}_{\text{mte}}(D_j, \bm{\Theta}')$
    
    \STATE \textbf{Domain Gradient Surgery}
    \STATE \quad Define $\mathcal{P}$ as $\{\bm{\Theta}\}$ if $M=\text{Global}$, else as $\{\bm{\Theta}_k\}$ for all layers
    \FOR{each element $p \in \mathcal{P}$}
        \STATE Extract corresponding gradients $\bm{G}_p$ and $\bm{F}_p$
        \IF{$\langle \bm{G}_p, \bm{F}_p \rangle < 0$}
            \STATE $\bm{G}_p \leftarrow \bm{G}_p - \frac{\langle \bm{G}_p, \bm{F}_p \rangle}{||\bm{F}_p||^2 + \epsilon} \bm{F}_p$
        \ENDIF
    \ENDFOR
    \STATE \textbf{Meta-optimization} $\bm{\Theta} \leftarrow \bm{\Theta} - \gamma (\bm{F} + \beta \bm{G})$
\ENDFOR
\end{algorithmic}
\end{algorithm}
This asymmetric strategy is crucial for stable meta-optimization, as it explicitly preserves the essential domain-specific knowledge provided by $\bm{F}$ for inner-loop adaptation, allowing $\bm{G}_{\text{proj}}$ to guide generalization effectively. The full procedure for base-DGS is integrated within our Algorithm \ref{alg:base_dgs} with the global mode.

\subsection{EMA-DGS: Persistent Anchor for Reliable Conflict Detection}
\label{subsec:ema_dgs}
The efficacy of DGS is fundamentally constrained by the stochastic volatility of instantaneous meta-train gradients ($\bm{F}$). Since $\bm{F}$ is derived from discrete mini-batches, it serves as a noisy estimator that varies substantially across training steps, even when the underlying optimization landscape is smooth. Such temporal instability triggers spurious conflict detections, where transient stochastic noise is misidentified as a genuine domain shift, leading to inconsistent projections that destabilize the bi-level optimization.

To resolve this problem, we propose EMA-DGS, which establishes a stable anchor by capturing the persistent optimization trajectory of the source domain (see Algorithm \ref{alg:ema_dgs} and Fig. \ref{fig:modeloverview}(c). Instead of relying on volatile per-batch signals, EMA-DGS maintains an exponential moving average (EMA) of $\bm{F}$ to decouple the underlying directional trend from noise. This refined direction serves as a temporally consistent reference that dominates the meta-test phase, ensuring that the surgery mechanism intervenes only in response to genuine domain-shift-induced conflicts, thereby improving the reliability and convergence of the meta-learning process.

\textbf{EMA of Meta-Train Gradient:} At each iteration $t$, after computing the current meta-train gradient $\bm{F}^{(t)}$, we update its EMA version $\bm{F}_{\text{EMA}}^{(t)}$ using a decay factor $\mu \in [0, 1)$:
\begin{equation}
\bm{F}_{\text{EMA}}^{(t)} = \mu \bm{F}_{\text{EMA}}^{(t-1)} + (1 - \mu) \bm{F}^{(t)}.
\label{eq:f_ema_update}
\end{equation}
For the first iteration, $\bm{F}_{\text{EMA}}^{(0)}$ is initialized to $\textbf{0}$. This $\bm{F}_{\text{EMA}}$ serves as a low-variance, more reliable representation of the historical trend of meta-train gradient directions, effectively filtering out high-frequency noise.

\textbf{Meta-Optimization with EMA-DGS:} EMA-DGS applies the same selective projection mechanism as the base-DGS (\ref{equ:base_dgs_proj}), but it uses the stable $\bm{F}_{\text{EMA}}$ as the anchor for detecting conflicts and projecting $\bm{G}$:
\begin{align}
\label{equ:ema_dgs_proj}
\bm{G}{\text{\small proj}} = 
\begin{cases}
    \bm{G} - \left(\frac{\langle \bm{G}, \bm{F}_{\text{EMA}} \rangle}{||\bm{F}_{\text{EMA}}||^2 + \epsilon} \right) \bm{F}_{\text{EMA}}, \quad &\text{if } \langle \bm{G}, \bm{F}_{\text{EMA}} \rangle < 0,\\
    \bm{G}, &\text{otherwise}.
\end{cases}
\end{align}
The model parameters are then updated using the current meta-train gradient $\bm{F}$ and the EMA-guided projected meta-test gradient $\bm{G}{\text{proj}}$:
\begin{equation}
\bm{\Theta} \leftarrow \bm{\Theta} - \gamma \left( \bm{F} + \beta \bm{G}_{\text{proj}} \right).
\label{eq:ema_dgs_update}
\end{equation}
By replacing the volatile instantaneous $\bm{F}$ with the stable $\bm{F}_{\text{EMA}}$ as the projection anchor, EMA-DGS promotes more consistent conflict resolution and a smoother trajectory.

\textbf{Why EMA Stabilizes the Anchor?}
The stabilization effect of EMA is grounded in its capacity for variance reduction \cite{kingma2014adam}. Assuming the instantaneous meta-train gradient is an unbiased estimator of the true gradient, $\bm{F}^{(t)} = \bar{\bm{F}} + \bm{\xi}^{(t)}$ with noise covariance $\sigma^2 \bm{I}$, the EMA recursion acts as a linear low-pass filter. In the stationary regime, the variance of the EMA anchor satisfies:
\begin{equation}
\text{Var}\left(\bm{F}_{\text{EMA}}^{(t)}\right) = \frac{1-\mu}{1+\mu}\sigma^2.
\label{eq:ema_var}
\end{equation}
The commonly used factor $\mu = 0.99$ yields a variance reduction factor of approximately $1/199$ relative to the raw gradient. 

Gradient noise suppression is critical for directional consistency. While the instantaneous anchor $\bm{F}^{(t)}$ fluctuates erratically due to independent noise $\bm{\xi}^{(t)}$, the EMA anchor integrates information over an effective window of $1/(1-\mu)$ steps. Consequently, EMA-DGS decouples the decision-making anchor from the parameter update: the stable $\bm{F}_{\text{EMA}}$ provides a reliable reference to identify genuine domain-shift conflicts, while the instantaneous $\bm{F}^{(t)}$ is preserved in the final update to maintain optimization responsiveness. This synergy prevents spurious interventions caused by transient stochasticity, leading to a more robust meta-optimization trajectory.


It is important to note that the EMA is applied \emph{only to the anchor} used for conflict detection and projection direction, not to the actual update gradient. The parameter update in (\ref{eq:ema_dgs_update}) still uses the current instantaneous $\bm{F}^{(t)}$, preserving full responsiveness to the latest gradient information. This asymmetric design, stable anchor for conflict detection combined with instantaneous gradient for update, avoids the double-smoothing effect that would arise if EMA were also applied to the update itself, which we empirically verify leads to performance degradation, especially combined with adaptive optimizers such as Adam.

\begin{algorithm}[t]
\caption{EMA-DGS-guided MLDG}
\label{alg:ema_dgs}
\begin{algorithmic}[1]
\STATE \textbf{Input} Source domains $\mathcal{S}$, decay $\mu$, hyperparameters $\alpha, \beta, \gamma$, Mode $M \in \{\text{Global, LW}\}$
\STATE \textbf{Init} Model parameters $\bm{\Theta}$, Meta-train EMA anchor $\bm{F}_{\text{EMA}} \leftarrow \bm{0}$
\FOR{each iteration}
    \STATE \textbf{Split} $\hat{\mathcal{S}}$ and $\tilde{\mathcal{S}} \leftarrow \mathcal{S}$
    \STATE \textbf{Meta-train} 
    \STATE \quad Compute average gradient 
     \STATE \quad $\bm{F} = \frac{1}{|\hat{\mathcal{S}}|} \sum_{D_i \in \hat{\mathcal{S}}} \nabla_{\bm \Theta}\mathcal{L}_{\text{mtr}}(D_i, \bm{\Theta})$
    \STATE \textbf{Update Anchor} $\bm{F}_{\text{ EMA}} \leftarrow \mu \bm{F}_{\text{EMA}} + (1-\mu) \bm{F}$
    \STATE \textbf{Inner-update} $\bm{\Theta}' \leftarrow \bm{\Theta} - \alpha \bm{F}$
    \STATE \textbf{Meta-test} 
     \STATE \quad Compute average gradient 
      \STATE \quad $\bm{G} = \frac{1}{|\tilde{\mathcal{S}}|} \sum_{D_j \in \tilde{\mathcal{S}}} \nabla_{\bm \Theta} \mathcal{L}_{\text{mte}}(D_j, \bm{\Theta}')$
    
    \STATE \textbf{EMA-Guided Gradient Surgery}
    \STATE \quad Define $\mathcal{P}$ as $\{\bm{\Theta}\}$ if $M=\text{Global}$, else as $\{\bm{\Theta}_k\}$ for all layers
    \FOR{each element $p \in \mathcal{P}$}
        \STATE Extract $\bm{G}_p$ and stable anchor $\bm{F}_{p, \text{EMA}}$
        \IF{$\langle \bm{G}_p, \bm{F}_{p, \text{EMA}} \rangle < 0$}
            \STATE $\bm{G}_p \leftarrow \bm{G}_p - \frac{\langle \bm{G}_p, \bm{F}_{p, \text{EMA}} \rangle}{||\bm{F}_{p, \text{EMA}}||^2 + \epsilon} \bm{F}_{p, \text{EMA}}$
        \ENDIF
    \ENDFOR
    \STATE \textbf{Meta-optimization} $\bm{\Theta} \leftarrow \bm{\Theta} - \gamma (\bm{F} + \beta \bm{G})$
\ENDFOR
\end{algorithmic}
\end{algorithm}

\subsection{LW-DGS: Fine-grained Gradient Surgery}
\label{subsec:lw_dgs}

Global gradient surgery projects gradients as single vectors, although conflicts may vary across the parameter tensors of large SSL models. Different layers encode distinct acoustic and contextual information \cite{el2025ssl}; consequently, a global projection may modify well-aligned layers while under-correcting conflict-prone ones.

Our previous layer-wise DGS (LW-DGS) \cite{qin2026dgsmldg} addresses this issue by applying gradient surgery independently to individual parameter tensors $\bm{\Theta}_k$, as shown by the LW mode in Algorithms~\ref{alg:base_dgs} and \ref{alg:ema_dgs}. Each tensor has corresponding meta-train and meta-test gradients, $\bm{F}_k$ and $\bm{G}_k$, with alignment measured by
\begin{equation}
\text{cos}(\bm{\Theta}_k) = \frac{\langle \bm{F}_k, \bm{G}_k \rangle}{||\bm{F}_k|| \cdot ||\bm{G}_k||}.
\end{equation}
A negative cosine similarity indicates a tensor-level conflict and triggers selective projection.

\textbf{LW-base-DGS:} Following our previous LW-DGS, when $\langle\bm{F}_k,\bm{G}_k\rangle<0$, we remove only the component of $\bm{G}_k$ opposing $\bm{F}_k$:
\begin{equation}
\label{eq:LW_proj}
\bm{G}_k^{\text{proj}} = 
\begin{cases}
\bm{G}_k - \frac{\langle \bm{G}_k, \bm{F}_k \rangle}{||\bm{F}_k||^2 + \epsilon} \bm{F}_k, & \text{if } \langle \bm{F}_k, \bm{G}_k \rangle < 0 \\
\bm{G}_k, & \text{otherwise}.
\end{cases}
\end{equation}
Here, $\epsilon$ ensures numerical stability. Otherwise, $\bm{G}_k$ remains unchanged, preserving aligned gradients.

\textbf{LW-EMA-DGS:} We further extend LW-DGS by using the EMA anchor $\bm{F}_{k,\text{EMA}}$ for each parameter tensor, rather than the instantaneous $\bm{F}_k$. The resulting tensor-specific projection is
\begin{equation}
\label{eq:LW_ema_proj}
\bm{G}_k^{\text{proj}} = 
\begin{cases}
\bm{G}_k - \frac{\langle \bm{G}_k, \bm{F}_{k, \text{EMA}} \rangle}{||\bm{F}_{k, \text{EMA}}||^2 + \epsilon} \bm{F}_{k, \text{EMA}}, & \text{if } \langle \bm{G}_k, \bm{F}_{k, \text{EMA}} \rangle < 0 \\
\bm{G}_k, & \text{otherwise}.
\end{cases}
\end{equation}
This variant combines the spatial selectivity of LW-DGS with EMA-based temporal stability. Conflict-prone tensors receive targeted corrections, while well-aligned tensors are left unchanged.

\section{Experimental Setup}
\label{sec:setup}

In this section, we detail the training sets, the data augmentation strategy, the backbone architectures under evaluation, the optimizer and hyperparameter settings, and the evaluation benchmarks and metrics.

\subsection{Datasets}
\label{subsec:datasets}
We conduct experiments under two training configurations to evaluate both the effectiveness and scalability of our framework.

\textbf{Single-dataset training.} We use the ASVspoof 2019 LA \cite{asvspoof2019} training set as the sole training source. It contains 6 spoofing attack types (A01--A06), each treated as an independent domain for meta-learning. This controlled setting isolates the effect of each proposed component.

\textbf{Multi-dataset training.} To evaluate scalability under a larger and more diverse domain pool, we additionally incorporate the ASVspoof 5 \cite{asvspoof5} and CFAD \cite{cfad2024} training sets. Both original training and development sets are employed. For meta-learning, each attack type is treated as an independent domain, yielding a total of 22 source domains: 6 from ASVspoof 2019 LA, 8 from ASVspoof 5, and 8 from CFAD. For ERM, all data are concatenated for joint training.

\subsection{Data Augmentation}
\label{subsec:augmentation}
To increase data variability in the single-dataset setting and ensure domain-shift emulation, we apply RawBoost \cite{rawboost} augmentation to the ASVspoof 2019 LA training set and treat the augmented copies as additional domains, yielding 12 source domains in total. Therefore, we consider noise variations in the experiment and also draw on insights from our case study. To separate the effect of data augmentation from that of episodic meta-learning, we include ERM-Aug as a control baseline. ERM-Aug is trained on the same original and RawBoost-augmented samples as the meta-learning methods, but pools all 12 domains and minimizes a single ERM objective without episodic splitting. In the multi-dataset setting, because the dataset configurations provide sufficient attack types and we consider the trade-off between training efficiency and data volume, we use only the augmented data from 22 attack types. This data augmentation is Rawboost-3.

RawBoost adds variation in the form of: i) linear and non-linear convolutive noise; ii) impulsive signal-dependent additive noise; iii) stationary signal-independent additive noise. In the code repository, Rawboost-3 applies stationary, signal-independent additive noise; Rawboost-5 applies linear and non-linear convolutive noise and impulsive, signal-dependent additive noise; and Rawboost-4 applies all three types of noise. full details are available in \cite{rawboost}. Augmentation with stationary, signal-independent additive noise is applied in the multi-dataset setting, where domain diversity is provided by the training datasets.

\subsection{Model Architecture}
\label{subsec:architecture}
We adopt the XLSR-Conformer-TCM \cite{truong24tcm} and XLSR-AASIST \cite{tak2022wav2vec} as our primary models for the single-dataset setting, and the XLSR-Mamba architecture \cite{xiao2025xlsr} for the multi-dataset setting. Both architectures share the same self-supervised XLS-R front-end (300M parameters) but differ in their backends, as shown in Figure \ref{fig:modeloverview}(b). Conformer-TCM combines self-attention with temporal convolutional modules, while BiMamba captures long-range temporal dependencies via selective state-space modeling.

\subsection{Training Configurations}
\label{subsec:training}
Audio segments are cropped or concatenated to approximately 4 seconds (64,600 samples at 16\,kHz). Training employs the Adam optimizer with a learning rate of $10^{-6}$ and weight decay of $10^{-4}$, optimizing the weighted cross-entropy loss. For the ERM training, the batch size is 20. For meta-learning, the batch size is 14 in the single-dataset setting and 12 in the multi-dataset setting, to fit the GPU memory during training. For the default configuration in meta-learning variants, the inner learning rate $\alpha$ is set to $10^{-3}$, and the meta-test weight $\beta$ is set to 0.5 based on validation performance. All models are trained with early stopping (patience 7) on NVIDIA RTX 4090 GPUs.

\subsection{Evaluation Settings}
\label{subsec:metrics}
\textbf{In-dataset evaluation.} For the single-dataset setting, we use the ASVspoof 2021 DF (\textbf{21DF}) test set \cite{yamagishi2021asvspoof}. For the multi-dataset setting, we additionally include the ASVspoof 5 (\textbf{ASV5}) test set \cite{asvspoof5}.

\textbf{Cross-dataset evaluation.} For the single-dataset setting, we use two benchmarks: \textbf{In-the-Wild (ITW)} \cite{muller2022itw}, a real-world dataset with uncontrolled recording conditions, and \textbf{Fake-or-Real (FoR)} \cite{reimao2019for}, which aggregates deepfake samples from various public sources. For the multi-dataset setting, we expand evaluation to five cross-dataset benchmarks: the CFAD unseen test set (\textbf{CFAD}) \cite{cfad2024}, \textbf{ADD 2023} \cite{yi2023add} test sets R1 and R2, \textbf{In-the-Wild} \cite{muller2022itw}, and \textbf{CodecFake} \cite{codecfake} (mean EER across 7 subsets).

\begin{table*}[t!]
\centering
\caption{Main results: EER (\%) across evaluation benchmarks. Bold and underline denote the best and second-best results per column. * denotes results reported in \cite{dowerah2026board}.}
\begin{tabular}{l|l|ccc|ccc}
\toprule
\multirow{2}{*}{\textbf{Framework}} & \multirow{2}{*}{\textbf{Training Paradigm}} & \multicolumn{3}{c|}{\textbf{XLSR-conformer-TCM \cite{truong24tcm}}} & \multicolumn{3}{c}{\textbf{XLSR-AASIST \cite{tak2022wav2vec}}} \\
\cmidrule(lr){3-5} \cmidrule(lr){6-8}
& & \textbf{21DF}$\downarrow$ & \textbf{ITW}$\downarrow$ & \textbf{FoR}$\downarrow$ & \textbf{21DF}$\downarrow$ & \textbf{ITW}$\downarrow$ & \textbf{FoR}$\downarrow$ \\
\midrule
\multirow{3}{*}{Baseline} & ERM \cite{truong2025addressing,tak2022wav2vec} & 2.06 & 7.79 & 10.69* & 3.69 & 10.46 & 7.47* \\
& ERM-Aug & 2.00 & 7.34 & 5.31 & 2.61 & 7.21 & 4.90 \\
& MLDG & 2.19 & 6.47 & 5.91 & 2.74 & 7.33 & 5.81 \\
\midrule
\multirow{4}{*}{CAML (Ours)} & base-DGS & 1.88 & 5.96 & 5.31 & 1.79 & 6.81 & 3.54 \\
& LW-base-DGS & 1.95 & 6.17 & 5.70 & 2.04 & 6.70 & 4.73 \\
& EMA-DGS & \underline{1.82} & \textbf{5.69} & \underline{4.82} & \textbf{1.43} & \textbf{4.98} & \textbf{2.87} \\
& LW-EMA-DGS & \textbf{1.57} & \underline{5.92} & \textbf{4.23} & \underline{1.73} & \underline{5.84} & \underline{3.41} \\
\bottomrule
\end{tabular} 
\label{tab:main_eer_comparison}
\end{table*}

\subsection{Performance Metrics}
We evaluate under both in-dataset and cross-dataset settings, reporting Equal Error Rate (EER, \%); lower values indicate better performance.

\section{Results and Discussion}
\label{sec:exp}

This section presents and interprets the experimental results in seven parts. Section~\ref{subsec:main_results} reports main results under the single-dataset setting, demonstrating the progressive gains of each CAML component. Section~\ref{subsec:sota} compares CAML against state-of-the-art systems. Section~\ref{subsec:multi_dataset} evaluates scalability under the 22-domain multi-dataset setting. Other sections provide ablation studies to isolate the contribution of key design choices and the hyperparameter sensitivity. Section~\ref{subsec:layer_wise_analysis} analyzes per-layer conflict patterns and the effect of EMA decay.

\subsection{Main Results}
\label{subsec:main_results}

Table~\ref{tab:main_eer_comparison} reports EER across three benchmarks for two SSL backbones, comparing three baseline paradigms and four CAML variants.

\myparagraph{Baselines} ERM shows substantial degradation on ITW and FoR, while ERM-Aug improves FoR EER from 10.69\% to 5.31\% for XLSR-Conformer-TCM and from 7.46\% to 4.90\% for XLSR-AASIST. However, standard MLDG does not consistently outperform ERM-Aug, indicating that domain-shift emulation alone is insufficient.

\myparagraph{Base-DGS} Applying asymmetric gradient surgery to MLDG improves performance on both backbones. For example, it reduces ITW EER from 6.47\% (MLDG) to 5.96\% on XLSR-Conformer-TCM and FoR EER from 5.81\% to 3.54\% on XLSR-AASIST, supporting the benefit of correcting conflicting meta-test gradients.

\myparagraph{EMA-DGS} Stabilizing the projection anchor further improves EER over base-DGS on all benchmarks for both backbones. The gains are especially clear on XLSR-AASIST, where EMA-DGS achieves 1.43\%, 4.98\%, and 2.87\% on 21DF, ITW, and FoR, respectively, illustrating the benefit of temporally stable conflict detection.

\myparagraph{LW-EMA-DGS} Tensor-level surgery yields the lowest EER on XLSR-Conformer-TCM for 21DF (1.57\%) and FoR (4.23\%). However, global EMA-DGS performs better on XLSR-AASIST, suggesting that the relative benefits of global and layer-wise projection vary across architectures.

\begin{table}[t]
  \centering
  \caption{Comparison of CAML with state-of-the-art SDD Systems (EER\%). * denotes results reported in \cite{dowerah2026board}.}
  \label{tab:sota_comparison_styled}
  \resizebox{\linewidth}{!}{%
    \begin{tabular}{l|l|ccc}
      \toprule
      \textbf{Framework} & \textbf{System / Variant} & \textbf{21DF} & \textbf{ITW} & \textbf{FoR} \\
      \midrule
      \multirow{9}{*}{Prior Work} & XLSR-AASIST \cite{tak2022wav2vec} & 3.69 & 10.46 & 7.47* \\
       & XLSR-MoE \cite{moe2025icassp} & 2.54 & 9.17 & - \\
       & XLSR-SLS \cite{zhang2024sls} & 1.92 & 7.46 & 5.07* \\
       & XLSR-Nes2Net-X \cite{liu2025nes2net} & 1.78 & 6.60 & 6.31* \\
       & XLSR-AASIST-SAM \cite{huang25e_interspeech} & 3.44 & 6.34 & 5.18 \\
       & Wav2DF-TSL \cite{hao2025moeijcnn} & 1.95 & 6.83 & - \\
       & DPDA-XLSR-conformer-TCM \cite{truong2025addressing} & 1.81 & 6.48 & 4.47 \\
       & DPDA-XLSR-Mamba \cite{truong2025addressing} & 1.74 & 6.43 & 4.86 \\
       & DPDA-XLSR-AASIST \cite{truong2025addressing} & 2.13 & 5.42 & 3.04 \\
      \midrule
      \multirow{3}{*}{Baseline 1} & XLSR-Conformer-TCM \cite{truong24tcm} & 2.06 & 7.79 & 10.68* \\
      &  \quad + ERM-Aug & 2.00 & 7.34 & 5.31 \\
       & \quad + MLDG & 2.19 & 6.47 & 5.91 \\
      \midrule
      \multirow{2}{*}{CAML (Ours)} & \quad +EMA-DGS & 1.82 & \textbf{5.69} & 4.82 \\
       & \quad +LW-EMA-DGS & \textbf{1.57} & 5.92 & \textbf{4.23} \\
       \midrule
       \multirow{3}{*}{Baseline 2} & XLSR-AASIST \cite{tak2022wav2vec} & 3.69 & 10.46 & 7.46* \\
      &  \quad + ERM-Aug & 2.61 & 7.21 & 4.90 \\
       & \quad + MLDG & 2.74 & 7.33 & 5.81 \\
      \midrule
      \multirow{2}{*}{CAML (Ours)} & \quad +EMA-DGS & \textbf{1.43} & \textbf{4.98} & \textbf{2.87} \\
       & \quad +LW-EMA-DGS & 1.73 & 5.84 & 3.41 \\
      \bottomrule
  \end{tabular}}
\end{table}

\subsection{Comparison with State-of-the-Art Systems}
\label{subsec:sota}

Table~\ref{tab:sota_comparison_styled} compares CAML against recent systems on the single-dataset benchmark. All systems use the XLSR 300M front-end, isolating the effect of training strategy.

Across both backbones, CAML consistently sets or ties the best result on every benchmark. 
With XLSR-AASIST, EMA-DGS surpasses the strongest prior system on this backbone, DPDA-XLSR-AASIST~\cite{truong2025addressing} (2.13\% / 5.42\% / 3.04\%), by margins on all three benchmarks. This comparison is particularly informative because DPDA also addresses gradient conflicts but uses symmetric PCGrad within an ERM-augmented pipeline, whereas CAML employs asymmetric surgery inside a meta-learning objective. The improvement suggests that asymmetry and bi-level structure together capture a strictly larger portion of domain-invariant signal than symmetric surgery alone.

XLSR-SLS~\cite{zhang2024sls} and XLSR-MoE~\cite{moe2025icassp} achieve their gains through architectural changes, selective layer stacking, and mixture-of-experts routing, respectively. CAML surpasses both without modifying the backbone, demonstrating that optimization-level intervention is competitive with architecture-level solutions and can, in principle, be combined with them for further gains.

\begin{figure}[t]
    \centering
    \includegraphics[width=\linewidth]{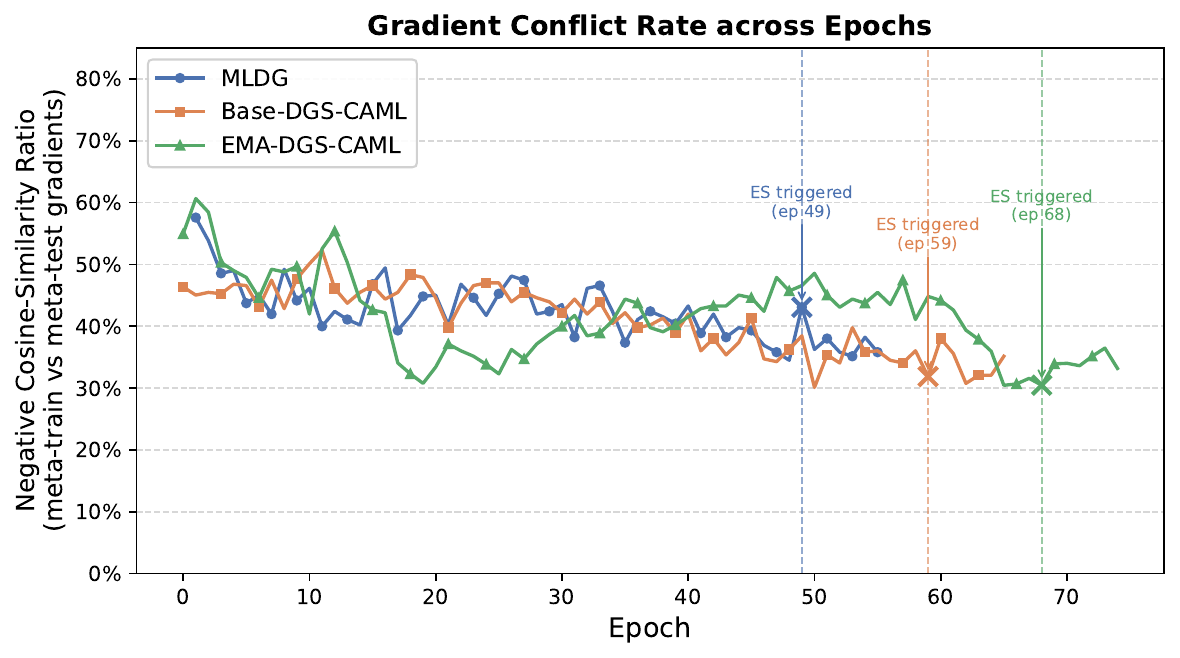}
    \caption{Gradient conflict rate (fraction of iterations where $\langle\bm{F},\bm{G}\rangle < 0$) across training epochs for MLDG, Base-DGS-CAML, and EMA-DGS-CAML. Vertical dashed lines mark the early-stopping (ES) trigger epoch for each method. The metric tracks the instantaneous meta-train gradient $\bm{F}$ against the meta-test gradient $\bm{G}$; the EMA anchor $\bm{F}_{\text{EMA}}$ is used only internally for projection and is not measured here.}
    \label{fig:coscom}
\end{figure}

\subsection{Gradient Dynamics Analysis}
\label{subsec:grad_dynamics}

This subsection examines whether CAML's asymmetric projection achieves its intended effect of resolving gradient conflicts during training, by tracking the cosine similarity between $\bm{F}$ and $\bm{G}$ across iterations or epochs.

\myparagraph{Optimization conflicts reduction}
To verify that CAML reduces gradient conflicts, we track the conflict rate, the fraction of training iterations in which $\langle\bm{F},\bm{G}\rangle < 0$, across epochs for MLDG, Base-DGS-CAML, and EMA-DGS-CAML. Note that $\bm{F}$ and $\bm{G}$ here are the instantaneous meta-train and meta-test gradients. As shown in Figure~\ref{fig:coscom}, all three methods begin with a high conflict rate ($\approx$45--60\%) at epoch~0, reflecting the frequent tension from effective domain-shift emulation. Both DGS variants exhibit a consistent downward trend, converging to $\approx$30\% by the time early stopping is triggered, whereas MLDG plateaus at a higher rate and stops considerably earlier.

A noteworthy observation is that EMA-DGS not only achieves the lowest conflict rate but also maintains a virtuous cycle of domain-shift emulation. We attribute this to a self-reinforcing dynamic: A stable projection anchor makes each gradient surgery decision more reliable, which allows the model to maintain effective domain-shift emulation for longer. This sustained emulation, in turn, gradually reduces the raw conflict between $\bm{F}$ and $\bm{G}$, thereby further improving domain-generalizable learning. The virtuous cycle that links anchor stability, domain-shift emulation, and conflict resolution. 

\begin{figure}[t]
    \centering
    \includegraphics[width=\linewidth]{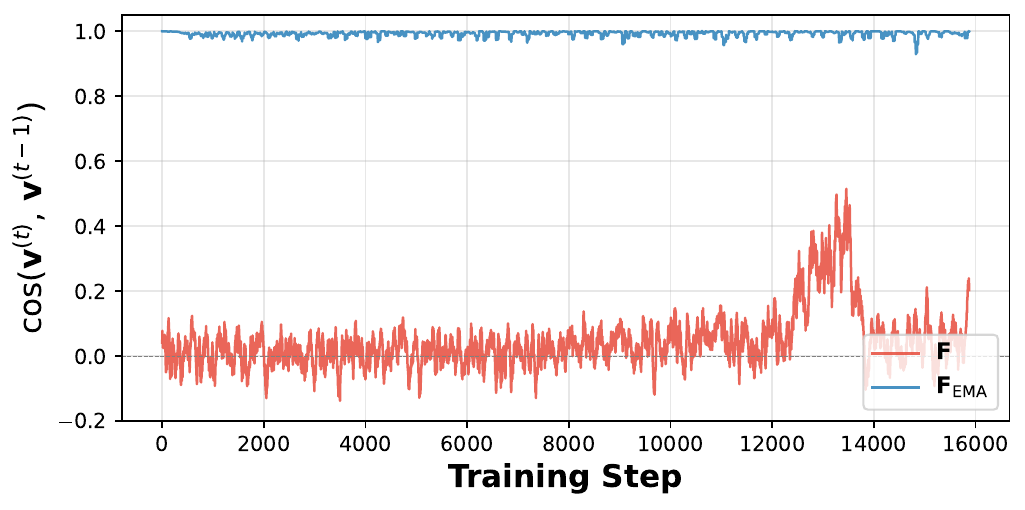}
    \caption{Direction consistency of the conflict detection anchor across consecutive training steps. The instantaneous $\bm{F}$ fluctuates significantly, while the EMA anchor $\bm{F}_{\text{EMA}}$ ($\mu = 0.99$) maintains high directional stability.}
    \label{fig:ema_stability}
\end{figure}

\myparagraph{EMA anchor stability}
To empirically validate the variance reduction analysis in Section \ref{subsec:ema_dgs}, we track the directional consistency of the anchor across consecutive training steps, measured by $\cos(\bm{v}^{(t)}, \bm{v}^{(t-1)})$ where $\bm{v}$ is the anchor used for conflict detection. As shown in Figure \ref{fig:ema_stability}, the instantaneous meta-train gradient $\bm{F}$ exhibits substantial directional fluctuations, with consecutive-step cosine similarity frequently dropping below 0.5 and occasionally becoming negative. This confirms that per-batch $\bm{F}$ is a volatile reference for conflict detection. Its erratic direction changes lead to inconsistent projection decisions across iterations. In contrast, the EMA anchor $\bm{F}_{\text{EMA}}$ maintains near-unity directional consistency throughout training, with $\cos(\bm{F}_{\text{EMA}}^{(t)}, \bm{F}_{\text{EMA}}^{(t-1)}) \approx 1$ for virtually all steps. This demonstrates that the exponential averaging effectively filters out high-frequency noise while preserving the underlying gradient trend, providing a stable and reliable reference for the DGS projection. The gap between the two curves directly illustrates the practical impact of the theoretical variance reduction factor derived in Equation \ref{eq:ema_var}.

\begin{figure}[t]
    \centering
    \includegraphics[width=0.95\linewidth]{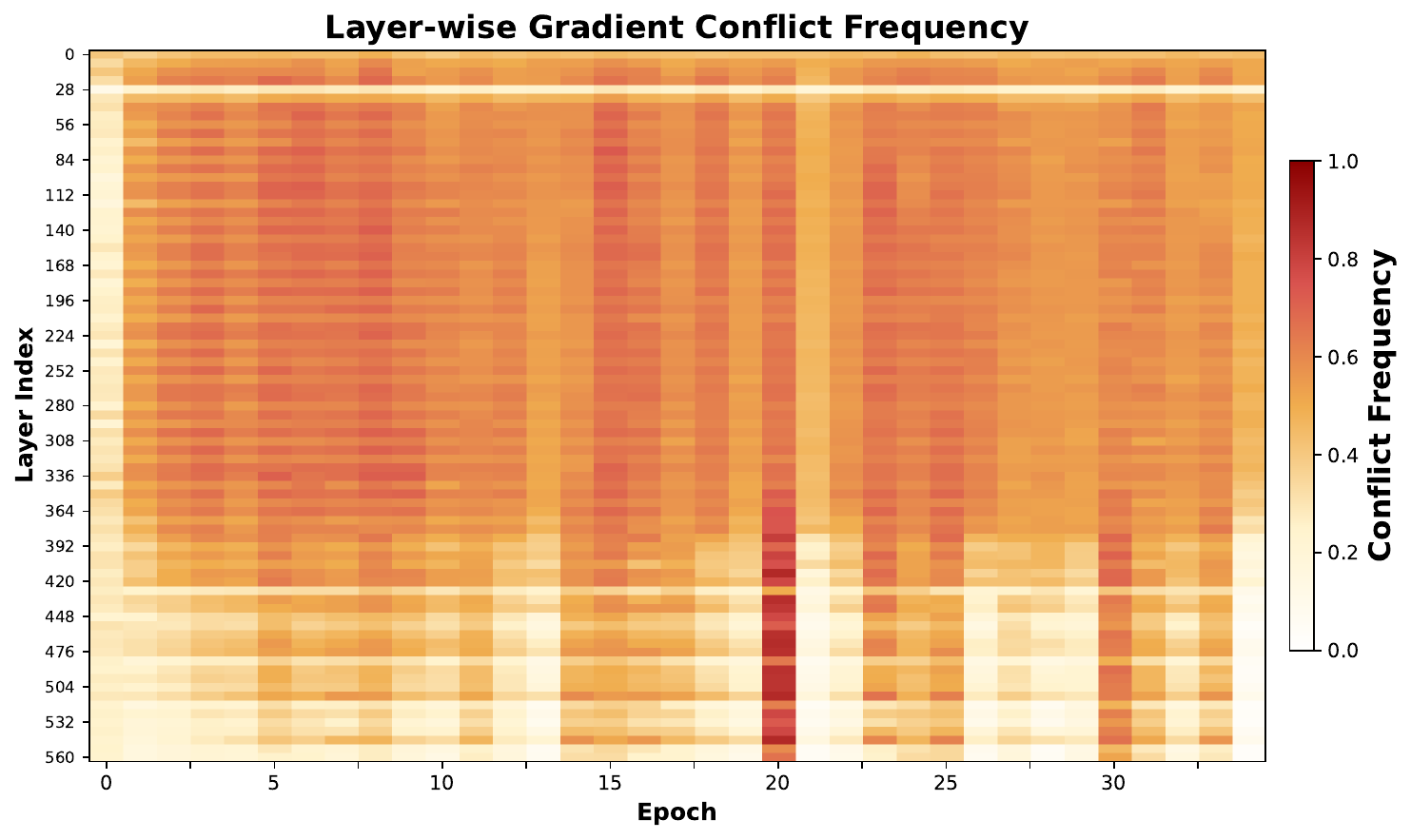}
    \caption{Layer-wise gradient conflict frequency across training epochs using LW-EMA-DGS. Each cell represents the fraction of steps within an epoch where the meta-train and meta-test gradients conflict at a given layer. Conflicts are heterogeneously distributed and evolve dynamically during training.}
    \label{fig:layer_conflict}
\end{figure}

\myparagraph{Layer-wise conflict distribution}
\label{subsec:layer_wise_analysis}
To justify the design of LW-DGS, we examine how gradient conflicts are distributed across network layers during training. Figure \ref{fig:layer_conflict} presents a heatmap of per-layer conflict frequency (the fraction of training steps in which $\langle \bm{F}_k, \bm{G}_k \rangle < 0$) across epochs. The heatmap covers the approximately 572 named-parameter tensors of the full model, which can be mapped to three functional groups based on the XLS-R 300M architecture: (1) the convolutional feature extractor, which computes a shallow mel-filterbank-like acoustic representation from the raw waveform; (2) the 24 stacked Transformer blocks of the SSL encoder, whose lower sub-layers extract sub-phonetic spectral envelopes and prosodic contours, while upper sub-layers encode speaker-level characteristics, recording-environment cues, and fine-grained naturalness patterns that neural vocoders and codec systems perturb; and (3) the downstream conformer-TCM \cite{truong24tcm} backend, which projects the contextualized speech representations onto a binary real/fake decision.

The results reveal two key observations aligned with this layer decomposition. (1) Spatially heterogeneous conflicts. Gradient conflicts are most severe in the upper SSL encoder, consistently exhibiting conflict frequencies above 0.6 throughout training. We reason that these layers encode speaker and recording-condition statistics that are precisely the domain-discriminative signals manipulated by different TTS/VC attack pipelines. The meta-train gradient $\bm{F}$ pushes these layers toward source-domain spoofing patterns, while the meta-test gradient $\bm{G}$ demands generalization across unseen synthesis strategies, producing structural directional tension. 
(2) Dynamic conflict evolution. The conflict distribution changes across training epochs. Some upper-level layers transition from moderate conflict in early epochs to low conflict by epoch 27, indicating increasing gradient alignment across source domains, while the lower layers continue to specialize. This dynamic behavior underscores the importance of per-step, per-layer conflict detection rather than a fixed intervention schedule.

\begin{table*}[t!]
\centering
\caption{Multi-dataset joint training results (\textbf{XLSR-Mamba} \cite{xiao2025xlsr}, 22 source domains): EER (\%) across evaluation benchmarks. Bold and underline denote the best and second-best results per column.}
\resizebox{\linewidth}{!}{
\begin{tabular}{l|l|cccccccc}
\toprule
\textbf{Framework} & \textbf{Variant} & \textbf{21DF}$\downarrow$ & \textbf{ASV5}$\downarrow$ & \textbf{CFAD}$\downarrow$ & \textbf{ADD2023 R1}$\downarrow$ & \textbf{ADD2023 R2}$\downarrow$ & \textbf{ITW}$\downarrow$ & \textbf{CodecFake}$\downarrow$ & \textbf{FoR}$\downarrow$ \\
\midrule
\multirow{2}{*}{Baseline} & ERM & 0.92 & \textbf{8.57} & 24.33 & 19.16 & 23.23 & 6.71 & 7.66 & 4.92  \\
 & MLDG & 1.05 & 9.69 & 24.16 & 18.14 & 23.03 & 6.47 & 7.24 & 4.60  \\
\midrule
\multirow{4}{*}{CAML (Ours)} & base-DGS & 1.00 & 10.24 & 23.86 & \textbf{17.76} & 23.19 & \underline{5.23} & \underline{6.76} & 4.55 \\
 & LW-base-DGS & 1.02 & 9.55 & \underline{23.83} & 17.88 & 22.68 & 6.15 & 7.27 & 4.34  \\
 & EMA-DGS & \textbf{0.87} & 9.41 & \textbf{23.79} & \underline{17.77} & \textbf{20.28} & \textbf{4.65} & 6.85 & \underline{4.19} \\
 & LW-EMA-DGS & \textbf{0.87} & \underline{9.14} & 24.10 & 17.95 & \underline{21.60} & 5.54 & \textbf{5.63} & \textbf{3.97}  \\
\bottomrule
\end{tabular}
}
\label{tab:multi_dataset}
\end{table*}

\begin{table*}[t]
\centering
\caption{Ablation study under the multi-dataset setting (XLSR-Mamba): EER (\%) across evaluation benchmarks.}
\resizebox{\linewidth}{!}{
\begin{tabular}{l|l|cccccccc}
\toprule
\textbf{Framework} & \textbf{Variant} & \textbf{21DF}$\downarrow$ & \textbf{ASV5}$\downarrow$ & \textbf{CFAD}$\downarrow$ & \textbf{ADD2023 R1}$\downarrow$ & \textbf{ADD2023 R2}$\downarrow$ & \textbf{ITW}$\downarrow$ & \textbf{CodecFake}$\downarrow$ & \textbf{FoR}$\downarrow$ \\
\midrule
\multirow{2}{*}{CAML (Ours)} & base-DGS & 1.00 & 10.24 & 23.86 & 17.76 & 23.19 & 5.23 & 6.76 & 4.55 \\
 & LW-base-DGS & 1.02 & 9.55 & 23.83 & 17.88 & 22.68 & 6.15 & 7.27 & 4.34 \\
\midrule
\multirow{2}{*}{Ablation} & \ding{172} Reverse Projection & 0.97 & 9.82 & 24.27 & 20.43 & 23.68 & 6.15 & 7.27 & 5.78 \\
 & \ding{173} SSL-Only LW-base-DGS & 1.16 & 9.57 & 23.93 & 18.03 & 22.75 & 6.58 & 7.27 & 5.91 \\
\bottomrule
\end{tabular}
}
\label{tab:ablation_eer_comparison}
\end{table*}


\subsection{Scalability Analysis}
\label{subsec:multi_dataset}

To assess scalability under greater domain diversity, we evaluate CAML using XLSR-Mamba \cite{xiao2025xlsr}, trained on 22 source domains from three datasets. Table~\ref{tab:multi_dataset} reports EER on eight benchmarks.

\myparagraph{EMA stabilization is the most effective component} EMA-DGS achieves the best or joint-best EER on 21DF (0.87\%), CFAD (23.79\%), ADD2023 R2 (20.28\%), and ITW (4.65\%) among all variants, supporting the benefit of a stable projection anchor in the multi-dataset setting.

\myparagraph{Layer-wise execution complements EMA} LW-EMA-DGS achieves the best EER on CodecFake (5.63\%) and FoR (3.97\%), and the best ASV5 result among CAML variants (9.14\%). By contrast, global EMA-DGS leads on ITW and ADD2023 R2. These complementary results suggest that layer-wise corrections may be preferable for certain domain shifts, while global projection better handles others. Both EMA variants improve over ERM and MLDG on most cross-dataset benchmarks.

\myparagraph{Base-DGS remains competitive} Without EMA stabilization, base-DGS still attains the lowest ADD2023 R1 EER (17.76\%) and competitive ITW (5.23\%) and CodecFake (6.76\%) performance, confirming the value of asymmetric projection.

\myparagraph{MLDG degradation on ASV5} All MLDG-based variants have higher EER than ERM on ASV5 (9.14--10.24\% vs.\ 8.57\%). This is consistent with the limited benefit of simulated domain shifts when evaluation conditions resemble training data, highlighting a trade-off between cross-domain robustness and performance on familiar distributions.

\begin{figure}[t]
    \centering
    \includegraphics[width=1.0\linewidth]{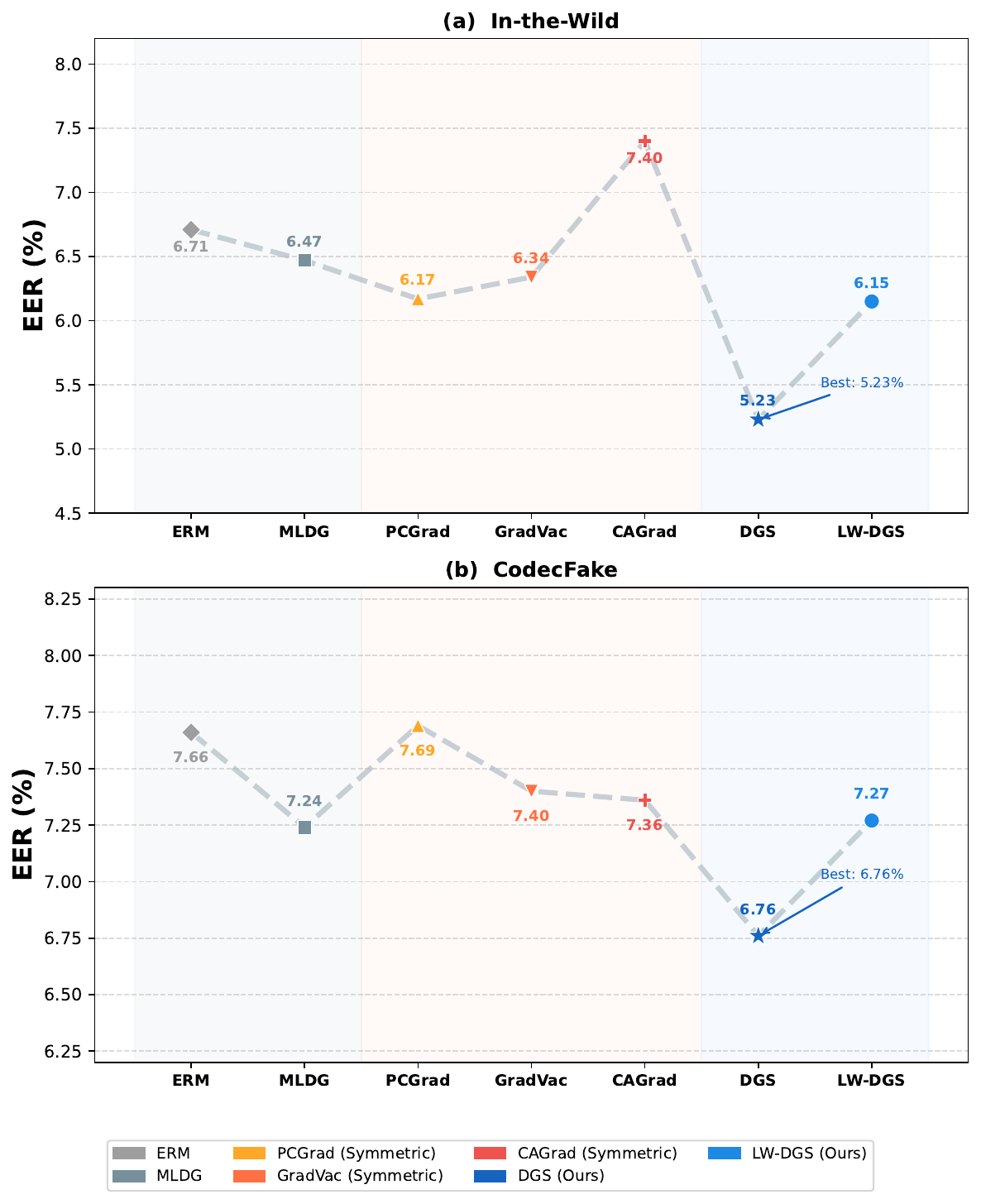}
    \caption{Comparison of symmetric and asymmetric gradient surgery methods on two cross-dataset benchmarks (EER\%) under the multi-dataset training setting (XLSR-Mamba, 22 source domains). }
    \label{fig:systemic}
\end{figure}

\subsection{Comparison with Symmetric Gradient Surgery Methods}
\label{subsec:symmetric}

To validate the necessity of the asymmetric design, we compare CAML against three representative symmetric gradient surgery methods from the multi-task learning literature, applied within the same MLDG framework.

Using the multi-dataset setting, Figure \ref{fig:systemic} compares our asymmetric DGS with symmetric multi-task gradient surgery methods: PCGrad \cite{yu2020gradient}, GradVac \cite{wang2020gradient}, and CAGrad \cite{liu2021conflict}. These methods treat the meta-train and meta-test objectives symmetrically by projecting both gradients in opposite directions.

Symmetric surgery does not consistently improve over standard MLDG. For instance, PCGrad achieves 6.17\% on In-the-Wild but degrades to 7.69\% on CodecFake (worse than MLDG's 7.24\%). CAGrad yields 7.40\% on In-the-Wild, also worse than MLDG. This is because symmetric projection modifies the meta-train gradient $\bm{F}$, which encodes the primary discriminative signal for source-domain adaptation. 

\subsection{Ablation Studies}
\label{subsec:ablation}

We conduct three ablation experiments to isolate the contribution of individual design choices in CAML. 
We conduct ablation experiments in the multi-dataset setting, as it provides broader benchmark coverage for validating design choices.

\subsubsection{Impact of Projection Direction}
To verify the necessity of asymmetry, we implement a reverse projection variant (Ablation \ding{172}) where $\bm{F}$ is projected onto the normal plane of $\bm{G}$, inverting the anchor and corrector roles. As shown in Table \ref{tab:ablation_eer_comparison}, this variant yields degradation on several benchmarks (20.43\% on ADD2023 R1, 23.68\% on ADD2023 R2), even underperforming the ERM baseline. This confirms that modifying the meta-train gradient $\bm{F}$ harms the model's ability to learn source-domain discriminative features.

\subsubsection{Effectiveness of Layer-Wise Monitoring}
Ablation \ding{173} restricts gradient surgery to the pre-trained SSL backbone only, leaving the downstream BiMamba backend untouched. Results in Table \ref{tab:ablation_eer_comparison} show higher EERs on most benchmarks compared to full LW-DGS (e.g., 18.03\% vs.\ 17.88\% on ADD2023 R1, 6.58\% vs.\ 6.15\% on In-the-Wild). This indicates that gradient conflicts are not confined to the feature extractor; the backend layers also exhibit domain sensitivity during adaptation. This validates the design of LW-DGS, which monitors and intervenes across all model parameters without imposing architectural constraints.

\begin{table}[t]
\centering
\caption{Effect of adding the $F_{\text{EMA}}$ to the meta-update. EER (\%) across evaluation benchmarks. $\mu = 0$ corresponds to base DGS without EMA.}

\begin{tabular}{lccc}
\toprule
\textbf{Decay $\mu$} & \textbf{21DF}$\downarrow$ & \textbf{ITW}$\downarrow$ & \textbf{FoR}$\downarrow$ \\
\midrule
$\mu = 0$ (base DGS) & 1.88 & 5.96 & 5.31 \\
$\mu = 0.99$ & \textbf{1.82} & \textbf{5.69} & \textbf{4.82} \\
\midrule
$\mu = 0.99$ ($\bm{F}_{\text{EMA}}$ to update) & 3.06 & 6.85 & 11.22 \\
\bottomrule
\end{tabular}
\label{tab:ema_update}
\end{table}

\subsubsection{Effect of Using \texorpdfstring{$F_{\text{EMA}}$}{FEMA} to The Meta-optimization}
The last row of Table \ref{tab:ema_update} validates the anchor-only design of EMA-DGS. When $\bm{F}_{\text{EMA}}$ is used not only as the projection anchor but also replaces the instantaneous $\bm{F}$ in the final parameter update (Equation \ref{eq:ema_dgs_update}), performance degrades severely across all benchmarks (3.06\%, 6.85\%, 11.22\% on 21DF, ITW, FoR respectively), substantially worse than even base DGS without EMA. This confirms the analysis in Section \ref{subsec:ema_dgs}: applying EMA to the update gradient introduces a double-smoothing effect with adaptive optimizers such as Adam, whose first-moment estimate already performs exponential averaging. The compounded smoothing over-dampens responsiveness to current gradient information, leading to stale updates and degraded convergence. This result underscores that the EMA should be applied exclusively to the conflict-detection anchor, while the actual parameter update must retain the instantaneous gradient to preserve the optimization dynamics.

\subsection{Analysis of hyperparameters}
\subsubsection{Effect of Inner-update Learning Rate \texorpdfstring{$\alpha$}{alpha}}
\label{subsubsec:ablation_alpha}

The inner-update learning rate $\alpha$ controls the step size of the inner-update (Equation~\ref{equ:mldg}), which determines how far the model is updated on the split source domains before testing on the meta-test domains.

\begin{figure}[t]
    \centering
    \includegraphics[width=\linewidth]{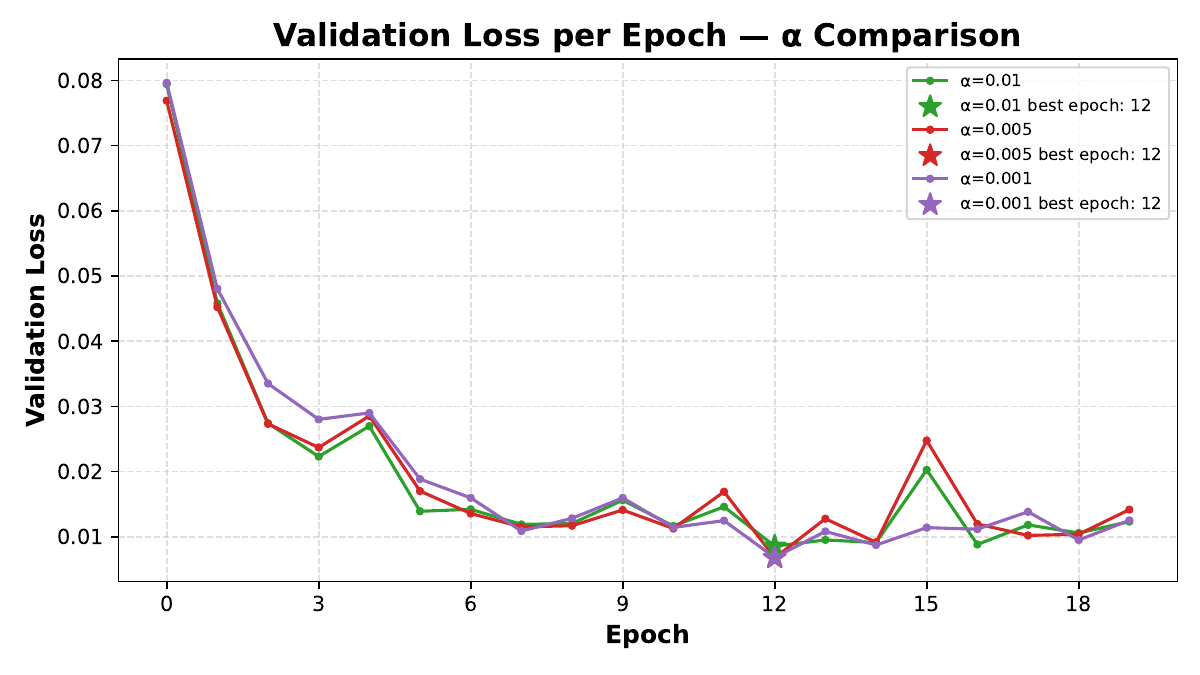}
    \caption{Validation loss curves of MLDG under different inner-update learning rates $\alpha$. Stars mark the best epoch for each setting (early stop trigger). Fixed $\beta$ is $0.5$ in this comparison.}
    \label{fig:alpha_compare}
\end{figure}

As shown in Figure~\ref{fig:alpha_compare}, validation loss curves across the range $\alpha \in \{0.01,\,0.005,\,0.001\}$ exhibit a clear optimum among the evaluated values.
The setting $\alpha = 0.001$ achieves the lowest final validation loss or the best balance between convergence speed and stability. Therefore, this value is adopted as the default across all experiments.

\subsubsection{Effect of Weighting Coefficient \texorpdfstring{$\beta$}{beta}}
\label{subsubsec:ablation_beta}

The weighting coefficient $\beta$ interpolates between the meta-train loss and the projected meta-test loss in the outer update (Equation~\ref{equ:mldg}).
When $\beta = 0$, the outer update reduces to standard supervised training on source domains; when $\beta = 1$, it places the same importance on the meta-test objective as the meta-train objectives
An optimal $\beta$ therefore reflects the relative importance of the two gradient signals after projection in Algorithm \ref{alg:base_dgs} and \ref{alg:ema_dgs}. 

\begin{figure}[t]
    \centering
    \includegraphics[width=\linewidth]{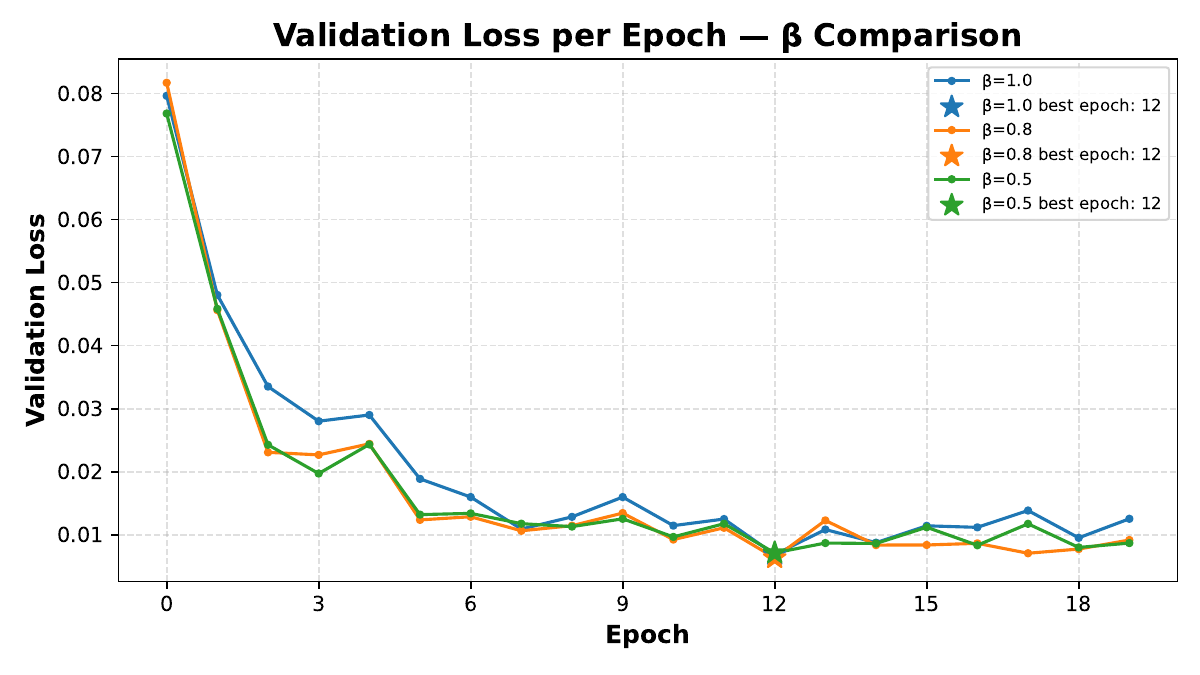}
    \caption{Validation loss curves of MLDG under different meta-update weighting coefficients $\beta$. Stars mark the best epoch for each setting. $\beta = 0.5$ achieves the most favorable convergence profile across all epochs. Fixed $\alpha$ is $0.001$.}
    \label{fig:beta_compare}
\end{figure}

Figure~\ref{fig:beta_compare} presents validation loss trajectories for $\beta \in \{0.5,\,0.8,\,1.0\}$.
Aggressive weighting ($\beta = 1.0$) yields the highest asymptotic loss, confirming that over-relying on the meta-test gradient.
At $\beta = 0.8$, the training curve improves but remains less stable after the early stop is triggered.
Setting $\beta = 0.5$ consistently produces the lowest validation loss across epochs and stability, striking an effective balance between source-domain learning and generalization pressure. This setting is used as the default in all reported experiments.

\subsubsection{Effect of EMA Decay Rate \texorpdfstring{$\mu$}{mu}}
Table~\ref{tab:ema_decay} reports EER and the absolute \texorpdfstring{$\Delta$}{Delta} relative to base-DGS (\texorpdfstring{$\mu=0$}{mu=0}) across all tested decay rates. Several observations emerge from the \texorpdfstring{$\Delta$}{Delta} columns.

\myparagraph{EMA consistently helps} Every tested $\mu > 0$ reduces EER on at least two of the three benchmarks compared to base-DGS, confirming the value of temporal smoothing. 

\myparagraph{A moderate window ($\mu = 0.77$--$0.90$) excels on 21DF (treating as the in-dataset setting)} The setting $\mu = 0.77$ achieves a notable 21DF reduction ($-$0.26\%), suggesting that a moderate smoothing window provides enough stability for detecting conflicts on in-dataset data while retaining sensitivity to recent gradient changes. The sole exception is 21DF at $\mu = 0.66$ (+0.05\%), which recovers immediately at $\mu \geq 0.77$.

\myparagraph{Long windows ($\mu \geq 0.99$) dominate on cross-dataset benchmarks (FoR and ITW)} $\mu = 0.999$ achieves the best EER ($-$1.30\%) on FoR, and the setting $\mu = 0.99$ achieves the largest ITW reduction ($-$0.27\%). A longer window yields an anchor that better represents the stable, global meta-train gradient direction, enabling more reliable conflict detection on distribution-shifted data.

\begin{table}[t]
\centering
\caption{Effect of EMA decay rate $\mu$ on EMA-DGS performance. EER (\%) and $\Delta$ (absolute change vs.\ base-DGS, lower is better). Bold denotes the best result per column; \colorbox{green!25}{green} / \colorbox{red!15}{red} cells indicate improvement / degradation $>$0.1\% relative to base-DGS.}
\setlength{\tabcolsep}{5pt}
\begin{tabular}{c|cc|cc|cc}
\toprule
\multirow{2}{*}{$\mu$} & \multicolumn{2}{c|}{\textbf{21DF}$\downarrow$} & \multicolumn{2}{c|}{\textbf{ITW}$\downarrow$} & \multicolumn{2}{c}{\textbf{FoR}$\downarrow$} \\
 & EER & $\Delta$ & EER & $\Delta$ & EER & $\Delta$ \\
\midrule
0 \small{(base-DGS)} & 1.88 & --- & 5.96 & --- & 5.31 & --- \\
\midrule
0.66  & \cellcolor{red!15}1.93  & +0.05 & 5.94             & $-$0.02 & \cellcolor{green!25}4.73 & $-$0.58 \\
0.77  & \cellcolor{green!25}1.62 & $-$0.26 & \cellcolor{green!25}5.74 & $-$0.22 & \cellcolor{green!25}4.99 & $-$0.32 \\
0.80  & \cellcolor{green!25}1.74 & $-$0.14 & 5.95             & $-$0.01 & \cellcolor{green!25}4.77 & $-$0.54 \\
0.88  & \cellcolor{green!25}1.69 & $-$0.19 & \cellcolor{red!15}6.09  & +0.13 & \cellcolor{green!25}4.34 & $-$0.97 \\
0.90  & \cellcolor{green!25}1.71 & $-$0.17 & 5.96  & +0.00 & \cellcolor{red!15}5.39  & +0.08 \\
0.99  & \cellcolor{green!25}1.82 & $-$0.06 & \cellcolor{green!25}\textbf{5.69} & $-$0.27 & \cellcolor{green!25}4.82 & $-$0.49 \\
0.999 & \cellcolor{green!25}1.78 & $-$0.10 & \cellcolor{green!25}5.79 & $-$0.17 & \cellcolor{green!25}\textbf{4.01} & $-$1.30 \\
\midrule
\rowcolor{gray!10}
0.99$^*$ & \textbf{1.57} & $-$0.31 & 5.92 & $-$0.04 & 4.23 & $-$1.08 \\
\bottomrule
\multicolumn{7}{l}{$^*$ LW-EMA-DGS (layer-wise execution, our best variant).}
\end{tabular}
\label{tab:ema_decay}
\end{table}

\myparagraph{Layer-wise execution amplifies the gains}
The bottom row shows that applying $\mu = 0.99$ in the layer-wise execution mode (LW-EMA-DGS) further reduces EER on 21DF to 1.57\% ($-$0.31\% vs. base-DGS) and FoR to 4.23\% ($-$1.08\%), confirming the complementary benefits of layer-wise conflict resolution and EMA stabilization.
Notably, $\mu = 0.99$ was selected based on the lowest validation loss, and the results demonstrate its effectiveness across all three benchmarks.

\section{Conclusion}
\label{sec:conc}
We have proposed the CAML, a conflict-aware meta-learning framework for domain-generalizable speech deepfake detection. We identify two optimization challenges in applying meta-learning for domain generalization (MLDG): gradient conflicts between meta-train and meta-test objectives, and the stochastic volatility of the conflict-detection anchor. CAML addresses both challenges through asymmetric domain gradient surgery and an exponential moving average anchor. The latter provides a temporally stable reference for reliable conflict detection. The layer-wise gradient surgery is proposed for fine-grained, targeted intervention. A key insight of this work is that applying an exponential moving average to the gradient anchor, rather than to model parameters, is an effective and principled approach to stabilizing the optimization dynamics for domain generalization in episodic training methods. CAML exhibits superior robustness and generalization, yielding significant performance gains across various backbones and achieving a relative reduction in error rates ranging from 21\% to 45\% compared to strong baselines.

Despite these improvements, several limitations remain. First, further evaluation with other self-supervised front-ends, broader acoustic and linguistic conditions, and multiple random seeds is needed to assess the generality and statistical robustness of the findings. 
Second, the improvements are not uniform across all conditions. On ASVspoof 5, the CAML variants yield higher EERs than ERM, indicating a possible trade-off between cross-domain robustness and performance on distributions close to the training data. Finally, the bi-level optimization requires separate meta-train and meta-test gradient computations, while the preferred EMA decay varies across benchmarks. Future work will investigate adaptive anchor selection with computationally efficient implementations.

\bibliographystyle{IEEEtran}
\bibliography{mybib}

\end{document}